# Twin boundary reversibility characteristics in α –Fe

J. Veerababu[1,2,*], G. Sainath[2] and A. Nagesha[1,2]

[1]Homi Bhabha National Institute, Mumbai, Maharashtra, India – 400 094

[2]Materials Development and Technology Division

Indira Gandhi Centre for Atomic Research, Kalpakkam – 603 102, India

[*]Corresponding author email: veeraj@igcar.gov.in

Phone: +91 44 27480118; Fax: +91–44–27480075

## Abstract:

Understanding the grain boundary deformation dynamics is very crucial to designing materials with stable microstructures. With this quest, the deformation behavior of coherent twin boundary under cyclic shear loading has been studied in α-Fe using molecular dynamics simulations to understand the influence of strain amplitude and temperature. Twin boundary exhibited shear coupled migration along with cyclic irreversibility character at lower temperatures and gained almost perfect reversibility at higher temperatures of around 1500 K. TB exhibited more sliding than migration with an increase in temperature. The stress associated with the migration of twins was observed to fluctuate around its average value. This was caused by layer by layer propagation of twin boundary through the activity of $1/6 \langle 111 \rangle = 1/12 \langle 111 \rangle + 1/12 \langle 111 \rangle$ type edge partial dislocations on $\{112\}$ twin plane. Complete twin reversibility and complex TB migration were noticed in the presence of multiple parallel twin nanowires. Twin migration was observed at the larger size nanowires with a proper combination of boundary conditions and strain rates; otherwise, it was absent. The influence of shear strain amplitude in offsetting the twin boundary was found to be minimal.

**Keywords:** Molecular dynamics, Twin boundary, Twin reversibility, Fe, Cyclic shear stress

## 1. INTRODUCTION

Reversible plasticity is an essential aspect of grain boundary engineering (GBE) to alleviate the damage accumulation in materials under cyclic loading. However, the practical methodology to achieve this technology in a vast majority of the materials is still challenging. Structural degradation and irreversible shear localization prevent deformation reversibility [1]. Several attempts have been made in the past to realize recoverable plasticity in bulk and nanoscale materials [2–7]. These interface engineering results showed that GB-mediated deformation helps in attaining reversible plasticity. Recently, Mohammadzadeh [8] reported reversible deformation in Fe–22Mn (wt.%) nanocrystalline TWIP steels by partial slip reversal and detwinning. Most recently, Zhu et al. [9] proposed the GBE method with controlled misorientation of low angle grain boundaries in FCC metallic bicrystals. They retain the plastic reversibility in the broad class of FCC metals under cyclic deformation. Yang Yang et al. [2] reported good stress-strain recoverability in pre-twinned α-Fe under torsion. Sainath et al. [10] reported coherent twin formation from incoherent twins and their stabilization in further cyclic loading in polycrystalline Cu. All these results underlined the crucial role of GBs in plastic reversibility.

Among the various GBs, coincidence site lattice (CSL) GBs are significant due to their peculiar influence on material properties [11]. Twin boundary (TB) which has the lowest sigma value (Σ3) with the lowest GB energy gains special attention among all other CSL GBs. It is the most frequently observed CSL GB in metals. TBs enhance strength without losing ductility in nanocrystalline materials [12]. Under the influence of stress, TBs can either deform [13] or migrate [14] depending on the material and loading conditions. Several experimental, theoretical, and molecular dynamics (MD) simulation studies have been performed to understand TB migration. Humberson and Holm [15,16] studied various mobility properties of incoherent twin boundaries in Ni using synthetic driving force in MD simulations. They reported the strong influence of boundary plane orientation of a GB on both the temperature dependence of mobility and the mechanisms of migration. Daphalapurkar et al. [17] reported that TB propagation encounters a drag force due to damping interaction with itself in addition to sharing common kinetic features of full dislocations. Wang et al. [18] studied incoherent TB (ITB) motion under shear in fcc metals. They related stacking fault energy with the disassociation of ITB into two phase boundaries and their further propagation. Generally, BCC metals show low twinning ability due to their high stacking fault energy [19]. However, alloying elements, temperature, strain rate and grain size change this criterion [20,21]. As a result, many technologically important BCC alloys undergo twinning deformation [20,22]. Ojha et al. [22] developed an analytical expression for twin nucleation stress in Fe and its alloys. Their theoretical work agreed well with the simulation and experimental results. They further proposed an overall stress expression to predict the twin migration stress in BCC systems [23]. Song et al. [24] reported irradiation-enhanced TB migration in BCC Fe using MD simulations. Unlike many other high angle GBs [25], TBs do not exhibit spontaneous migration or sliding at high temperatures [26].

All the above studies indicate the migration character of TB under uniaxial loading. In this context, it is important to understand the TB motion under cyclic loading. In the past, reversible twin boundary migration has been reported in HCP/FCC and BCC nanowires [2,27–29]. However, the reversibility of TBs under cyclic loading in BCC structures is rarely investigated. An attempt has been made in this study to characterize the TB reversibility in BCC Fe under cyclic shear loading.

## 2. SIMULATION DETAILS

Large-scale Atomic/Molecular Massively Parallel Simulator (LAMMPS) package [30] has been used for MD simulations. Mendelev et al. [31] embedded atom method (EAM) potential has been used to calculate inter-atomic forces between Fe atoms. This potential is highly reliable and widely used in literature to study plastic deformation in BCC Fe [13]. A simulation box with dimensions of $21 \times 6 \times 11.8$ nm$^3$, containing 1,29,600 Fe atoms has been chosen for the present study (Fig. 1a). The simulation box has an aspect ratio (length (L)/width (W)) of 1.77. Coherent twist twin boundary was introduced in the simulation box at the center of the X [121] - axis by rotating one part of the crystal relative to the other part by 180° around the <112> axis. This twin boundary ($\Sigma 3(112)$) in α-Fe does not carry the perfect mirror symmetry [32]. Periodic boundary conditions were applied in the Y [-101] direction, while the other two directions (X [121], Z [1-11]) were free to mimic the free surfaces. The simulation box has been divided into three regions, *viz*, two rigid regions at the top and the bottom portions along the X-axis, and the remaining middle portion as the active deformation region. The rigid bottom end was kept fixed, and cyclic shear load was applied on the rigid top end as shown in Fig. 1(a).

The initial stable structure was obtained through the conjugate gradient energy minimization method. This stable structure was further thermally equilibrated for a duration of 40 ps in NVT ensemble under zero applied force on rigid boundaries at fixed temperature using the Nose-Hoover thermostat. Initial velocities to the atoms were assigned randomly from finite temperature Maxwell distribution. The equilibrated model system is shown in Fig. 1(b). After equilibration, deformation simulations were carried out with a constant strain rate $5 \times 10^8$ /s at different temperatures ranging from 10-1500 K under NVT ensemble with a velocity verlet algorithm to integrate the Newton equations of motion with a time step of 1 fs. Strain-controlled shear cyclic loading simulations were employed on the model system under fully reversed triangular waveform as shown in Fig. 1(a). Shear strain is defined as below:

$$\text{Shear strain} = \frac{\Delta L_Z}{L_X} \tag{1}$$

The virial stress definition was used to calculate the stress values. AtomEye [33] and OVITO [34] packages were used to visualize the atomic configurations. Centro-symmetry parameter (CSP) coloring methodology was assigned in AtomEye to visualize the defects and atomic rearrangements. The simulation matrix employed in the present study is listed in Table 1. Conditions mentioned against serial numbers 1 to 3 and 7 to 9 were used to study the effect of strain amplitude (displacement), while all the test conditions were used to study the effect of temperature on twin boundary motion.

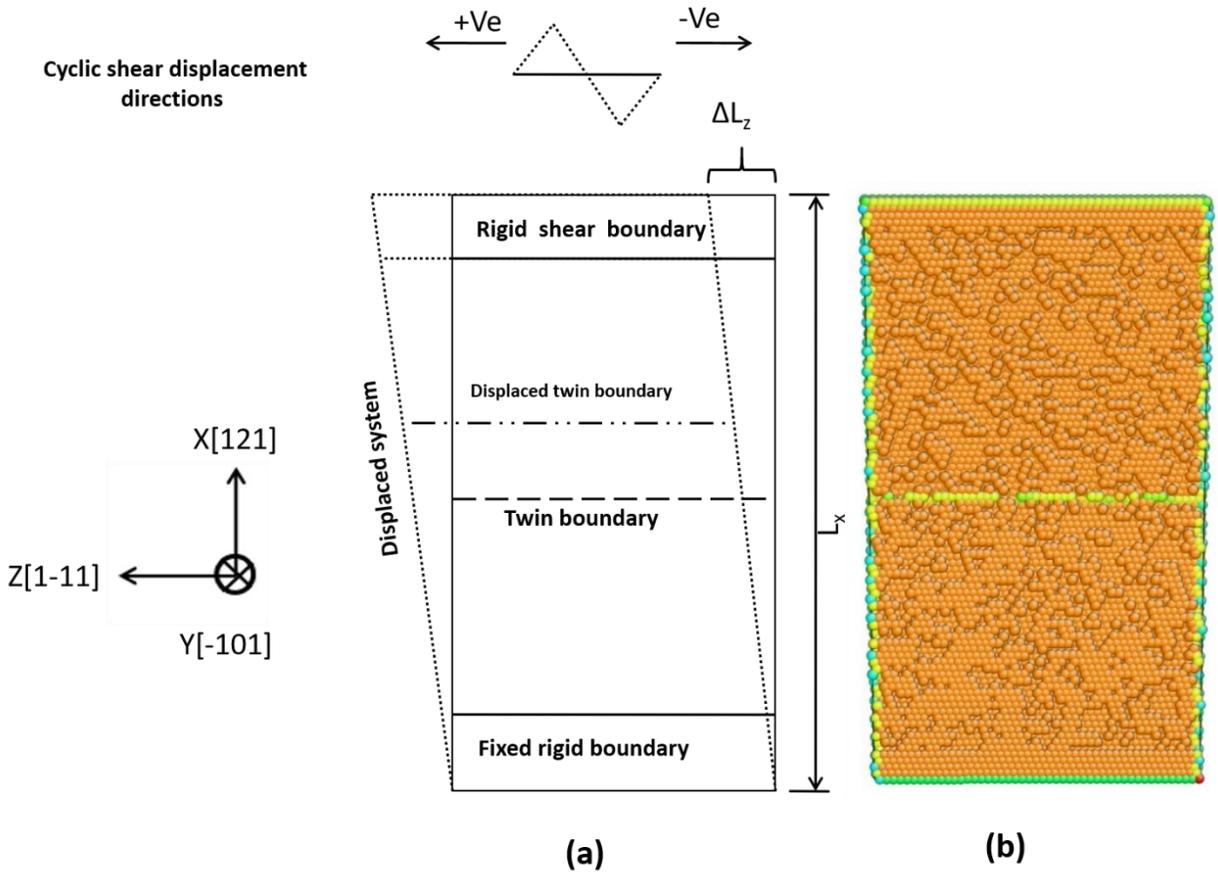

Fig. 1 (a) Schematic illustration of the model system under shear loading before shear displacement (solid lines) and after shear displacement (dotted lines). Twin boundary ($\Sigma3(112)$) displacement is marked by the dash-dot-dot line. (b) The equilibrated model system with twin boundary at the center.

Table 1: Simulations matrix employed in the study.

| S.N. | Temperature (K) | Shear strain amplitude (%) |
|------|-----------------|----------------------------|
| 1 | 10 | 5.86 |
| 2 | 10 | 14.76 |
| 3 | 10 | 24.63 |
| 4 | 100 | 14.76 |
| 5 | 300 | 14.76 |
| 6 | 600 | 14.76 |
| 7 | 800 | 5.86 |
| 8 | 800 | 14.76 |
| 9 | 800 | 24.63 |
| 10 | 1000 | 14.76 |
| 11 | 1500 | 14.76 |

## 3. RESULTS

### 3.1 Twin boundary migration and reversibility

The atomic snapshots of twin boundary displacements at different strain levels under shear loading on the model system at 10 K are shown in Figs. 2(A-E). These atomic structures indicate the migration of the twin boundary from its original position (dashed line). As shown in Fig. 2 (B), the TB displaced towards the top rigid region as the applied strain reached its peak value of 14.76%. One can also observe the formation of the step at the TB upon migration. This type of behavior is called shear coupled migration (SCM). The TB reversed its migration upon changing the strain direction and exhibited an offset at the zero applied strain as shown in Fig. 2 (C). The TB moved towards the rigid bottom region with further straining in the reverse direction (Fig. 2 (D)). Again, reversing the strain direction caused a change in the direction of migration of the TB, exhibiting an offset at the zero strain (Fig. 2 (E)). Thus, the system was finally left with an offset value in the TB displacement from its original position after a complete shear cycle. One can also observe the symmetrical SCM of the TB along with the forward and backward applied shear directions (Fig. 2 (B)- (E)). A low offset value indicates more recovered plasticity and vice versa. Thus, the twin boundary has an irreversible SCM character under cyclic shear loading.

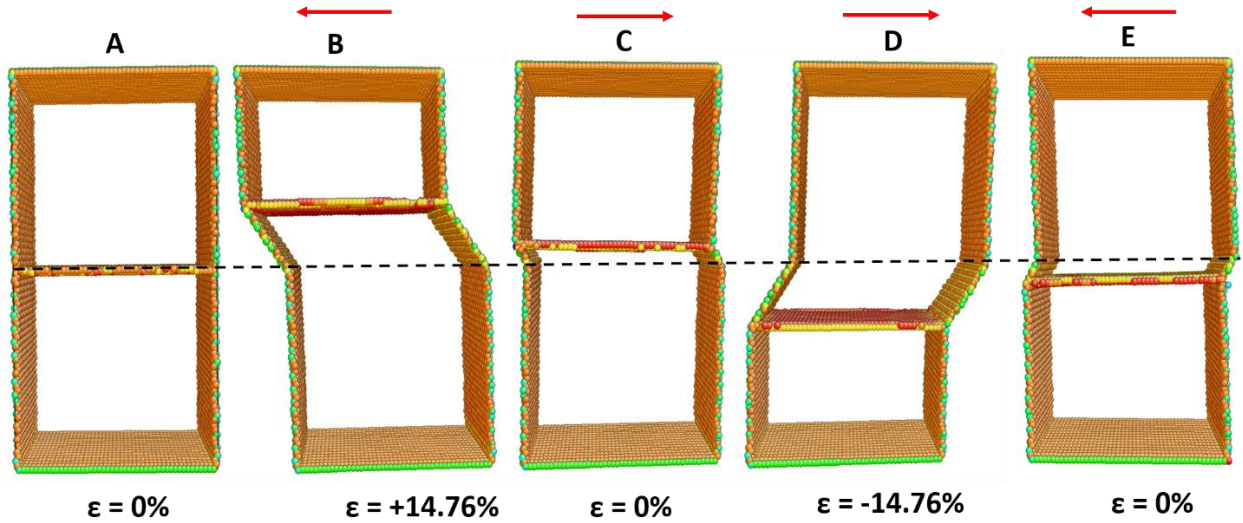

Fig. 2 SCM of TB at different cyclic strains at a temperature of 10 K. A: initial position of the TB, B: the migrated TB position after the application of forwarding stress, C: the position of the TB after removing the forwarded stress, D: migrated TB position after the application of reverse stress, E: position of the TB after a complete shear cycle. C & E indicate the offset in TB with reference to the initial TB position in A as indicated by the dashed black line. For clarity, perfect BCC atoms are removed. Centro symmetric coloring is assigned.

The mechanism operating at the atomic level to displace the twin boundary under applied shear stress can be understood from Fig. 3. Application of external shear stress along Z ⟨1 1 1⟩ direction leads to the formation of 1/6 ⟨1 1 1⟩ = 1/12 ⟨1 1 1⟩ + 1/12 ⟨1 1 1⟩ type edge partial dislocations on adjacent {112} twin boundaries as shown in Fig. 3 (i) & (ii). These edge partials move on twin planes in opposite directions and annihilate at opposite surfaces (as shown in Fig. 3 (ii) – (iv)). This process removes one atomic layer on the back end of the twin plane and adds a new layer at the front end (moving direction), eventually resulting in the growth of the twin plane by one atomic layer ahead in the X ⟨1 1 2⟩ direction. The formation of dislocation loops was also observed in this process as shown in Fig. 3 (v). Repeated operation of this mechanism leads the twin plane to move layer by layer until strain reversal. Upon strain reversal, the same mechanism operates on the twin plane to displace it in the reverse direction.

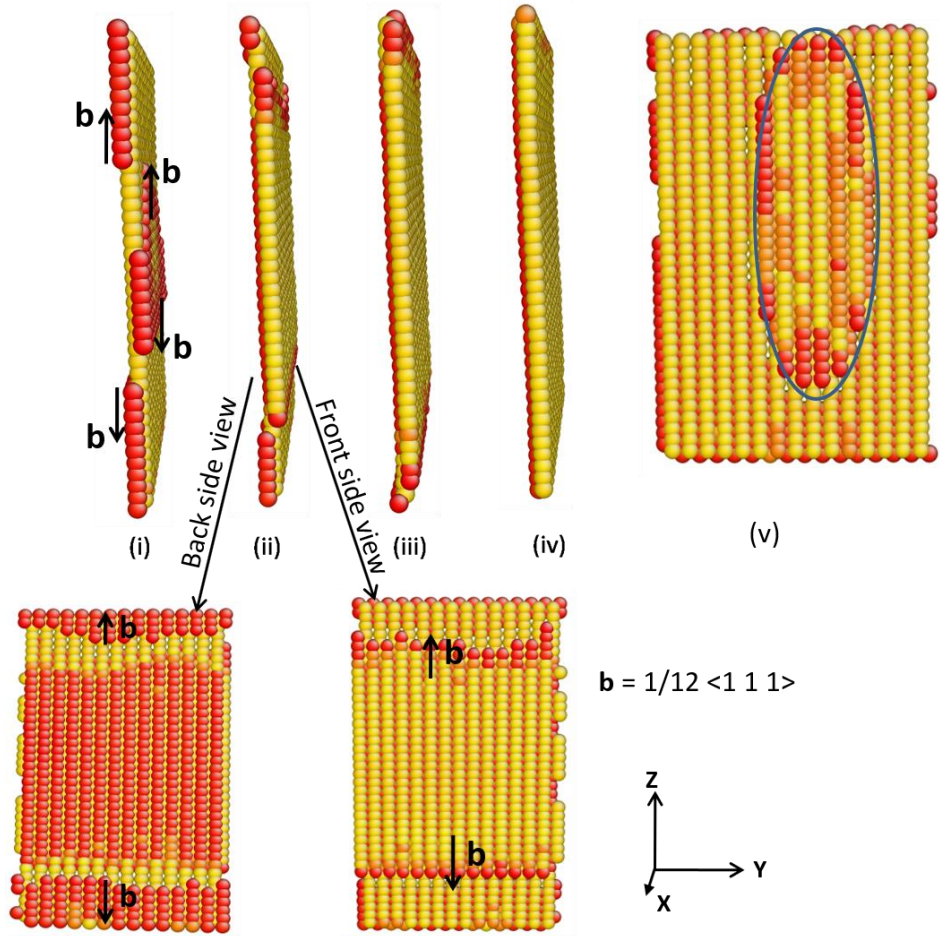

Fig. 3 The mechanism of twin plane displacement (i) formation of 1/12 ⟨1 1 1⟩ type edge partials (ii) back and front side views of twin plane and directions of burger vectors of the partial dislocations (iii) propagation of partial dislocations towards opposite surfaces (iv) annihilation of dislocations at the respective surfaces (v) observed dislocation loop during twin propagation.

### 3.2 Strain amplitude effect

The SCM characteristics have been plotted in Fig. 4 for an applied cyclic shear strain amplitude of ±14.76 % at 10 K for the first two cycles. Further, the respective locations of the atomic structures have shown in Fig. 4 can be identified from Fig. 2 (A to E.) The displacement of the twin boundary was calculated by taking the average of all the atom positions on the twin plane. The twin boundary was cyclically displaced under shear strain. Initially, the twin boundary was not moving upon the application of shear strain (location A in Fig. 4) till a critical strain ($\varepsilon_{critical}$) of 4.38% was attained (location X in Fig. 4). Further, this kind of inertia (lagging) of the twin boundary could be observed from Fig. 4 at each reversal of the shear strain. The lagging time ($t_{lag}$), defined as the time taken by the model system to initiate the movement of twin boundary or reversal of the direction of the twin boundary at the peak strains, is the same in all forward or all backward peak strains (at locations Y and Z in Fig. 4) and further $t_{lag}^{Backward} = 2\ t_{lag}^{Forward}$. However, one can notice the identical dynamics of the twin boundary for the first two cycles. Variation in the offset value (at the completion of a cycle) with the applied shear strain amplitude is presented in Fig. 5. A larger strain amplitude could cause a more significant displacement of the TB. Still, the offset value was constant at both the temperatures 10 K and 800 K. Thus, the effect of strain amplitude on the offset value was found to be negligible.

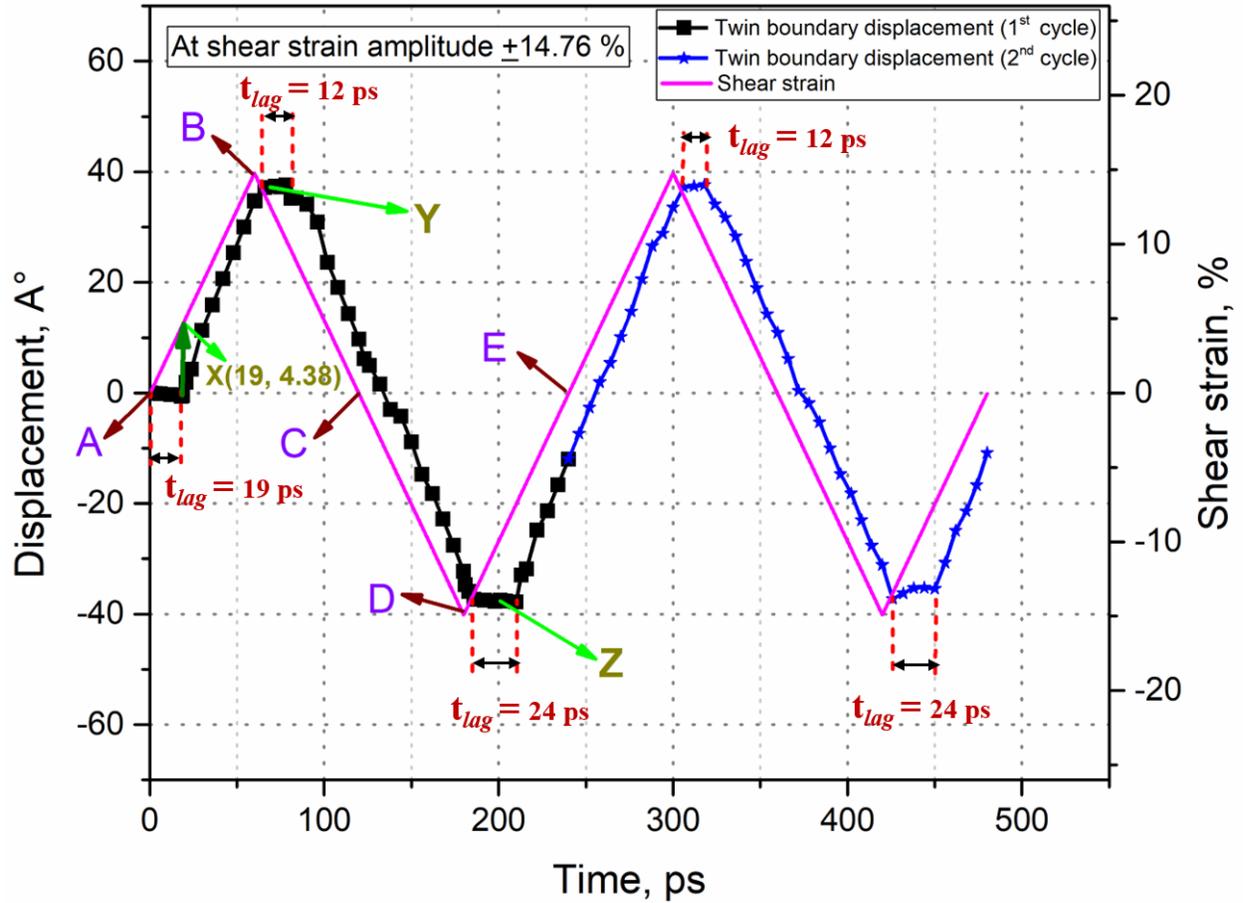

Fig. 4 Cyclic nature of twin boundary motion at 10 K for a shear strain amplitude of ±14.76% and microstructures corresponding to the indicated locations A, B, C, D & E are shown in Fig. 2(A-E) respectively.

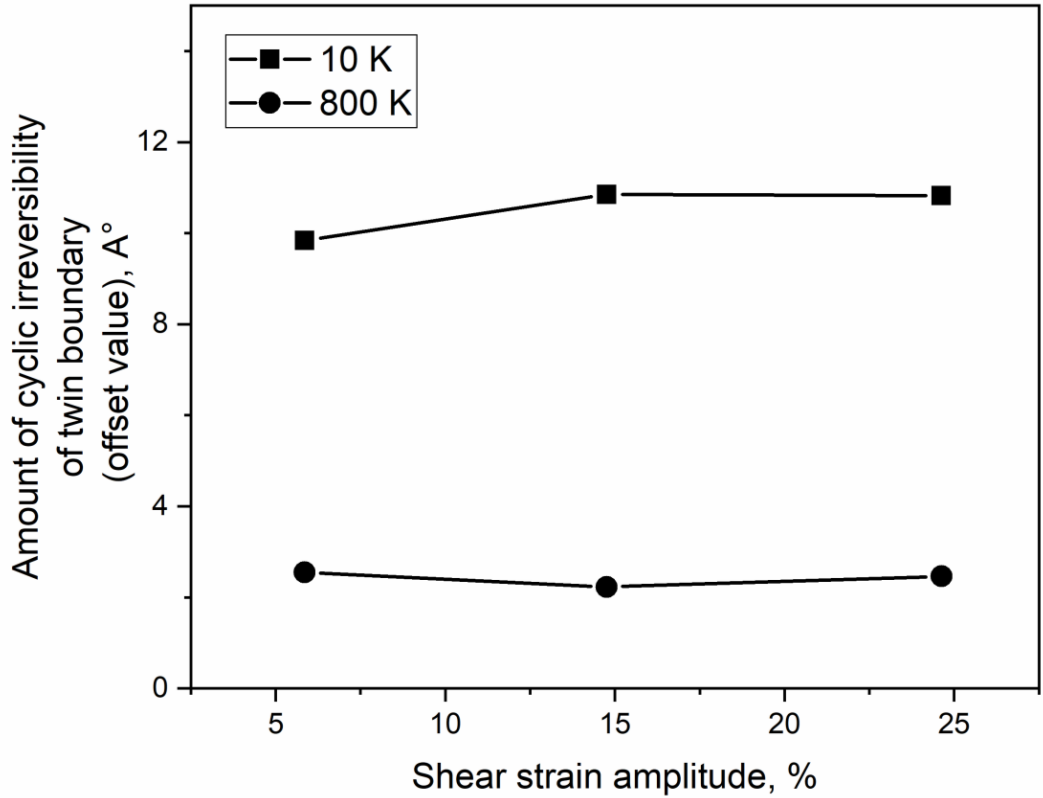

Fig. 5 Change in offset value after completion of a cycle with the applied cyclic shear strain amplitude at the temperatures 10 K and 800 K.

### 3.3 Temperature effect

The effect of temperature on the twin boundary reversibility was investigated at a fixed shear strain amplitude of ±14.76%. Figure 6 shows the displacement characteristics of the twin boundary at different temperatures ranging from 10 to 1500 K. Initial lagging during strain reversals of the twin boundary was observed to diminish gradually with increasing temperature. Increasing fluctuations in the twin boundary displacement with the increase in temperature can be evident from Fig. 6. The effect of temperature on the offset value of the twin boundary is shown in Fig. 7. The offset value followed an exponential decay behavior and reached close to zero at 1500 K. Twin boundary in the model system at 1500 K is shown in Fig. 8. It can be understood from the above figure that a further increase in the temperature results in a not so well-defined twin boundary. Thus, at higher temperatures, the twin boundary attained an almost reversible character. It is important to notice that the melting temperature of Fe is 1811 K and it undergoes a phase transition from BCC to FCC at 1183 K and again FCC to BCC at 1665 K, which the present potential cannot predict. However, the observed trend clearly demonstrates that the offset value decreases with increasing temperature beyond 1183 K, notwithstanding the above limitation.

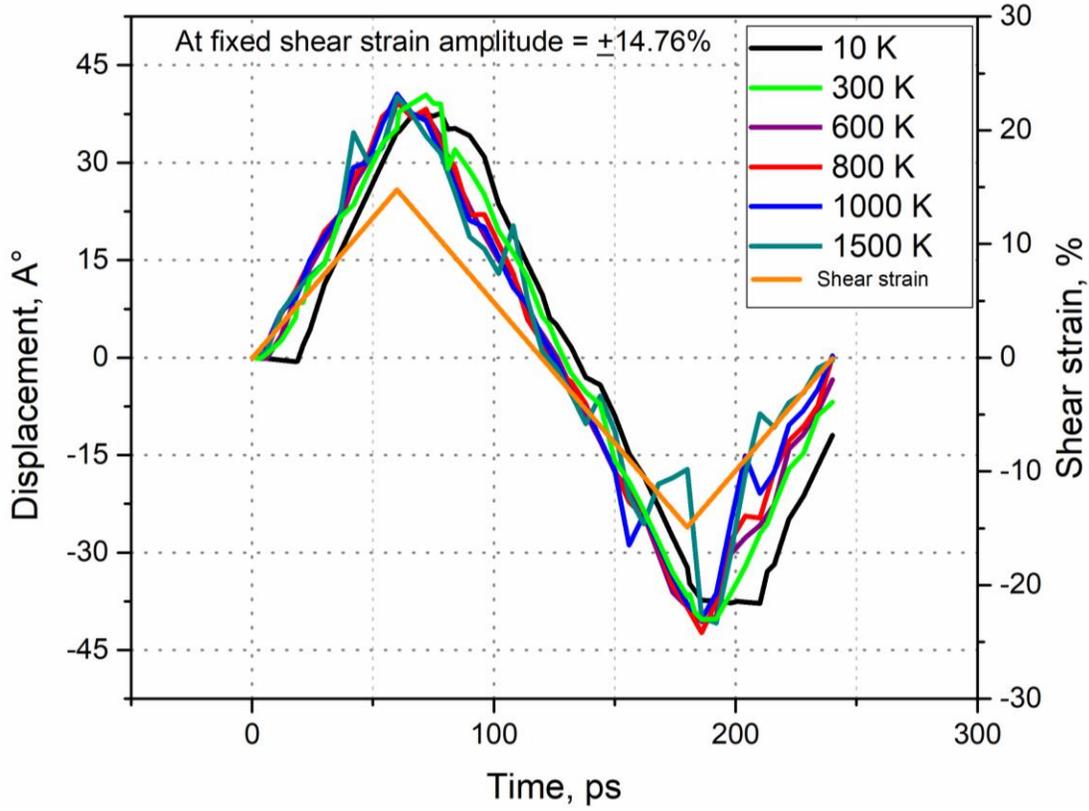

Fig. 6 Effect of temperature on twin boundary displacement concerning applied shear strain for the first cycle.

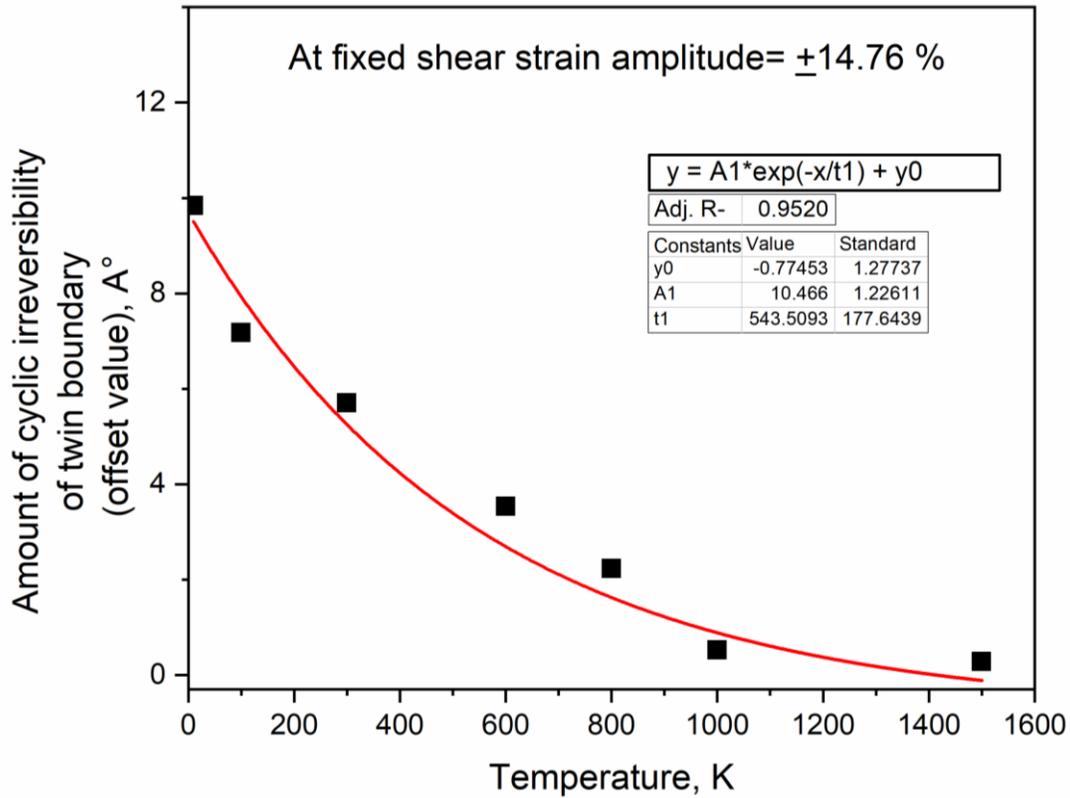

Fig. 7 Variation of TB offset value (after completion of a cycle) with temperature. Data are fitted with an exponential decay curve as indicated in the figure.

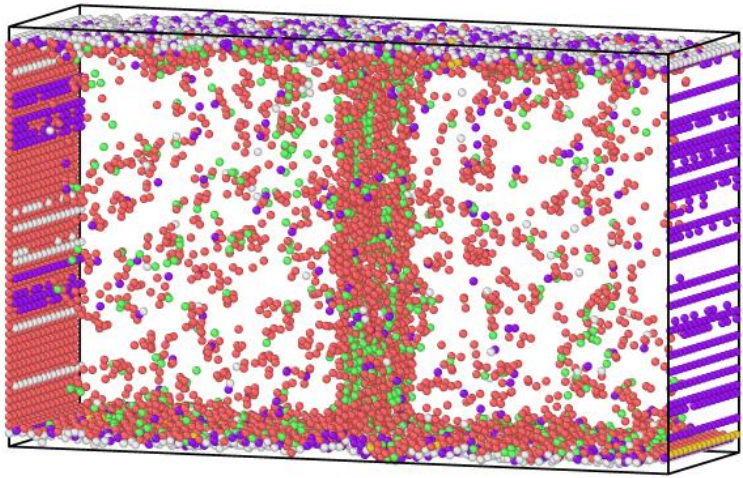

Fig. 8 Model system at 1500 K (perfect BCC atoms have been removed for clarity). Numerous point defects are evident in the system.

Figure 9 depicts the cyclic stress response of the system for the first cycle at different temperatures along with the input shear strain waveform. Stress values for a given temperature initially increased linearly until the system reached a peak point (e.g., location $P$ at 10 K). This indicates the elastic deformation in the system. Stress drop at location $P$ represents the initiation of twin boundary motion. Further, this stress drop had settled down to a saturated average value (e.g., 1.2 GPa at 10K), denoted as the twin migration stress (TMS). This TMS value is the same at all the peak strains in Fig. 9 (negative sides & positive sides) for a given temperature. However, the TMS value decreased exponentially with increasing temperature, as shown in Fig. 10. This TMS either can induce TB motion normal to the boundary or trigger TB sliding. Hence TB attains two types of velocities namely velocity normal to TB (normal/migration velocity, $v_n$) and velocity parallel to TB (sliding velocity, $v_s$). The geometric coupling factor [25,26], defined in equation (2) can be used further to study the behavior of TB migration:

$$\beta = \frac{v_{||}}{v_n} \tag{2}$$

where, $v_{||}$= top rigid boundary velocity. Table 2 listed out the values for $\beta, v_s$ at different temperatures. Reduction in $\beta$ value with increasing temperature indicates the weakening of the coupling between applied velocity and migration velocity. An increase in temperature reduced the migration velocity but increased the sliding velocity, indicating the reduction in the resistance offered by the adjacent planes to the TB.

Table 2 $\beta$, $v_s$ values at different temperatures.

| Temperature (K) | $\beta$ | $v_s$ |
|---|---|---|
| 10 | 0.1241 | 0.4288 |
| 100 | 0.1343 | 0.4711 |
| 300 | 0.1345 | 0.4803 |
| 600 | 0.1351 | 0.4961 |
| 800 | 0.1455 | 0.5079 |
| 1000 | 0.1475 | 0.5084 |
| 1500 | 0.1596 | 0.5105 |

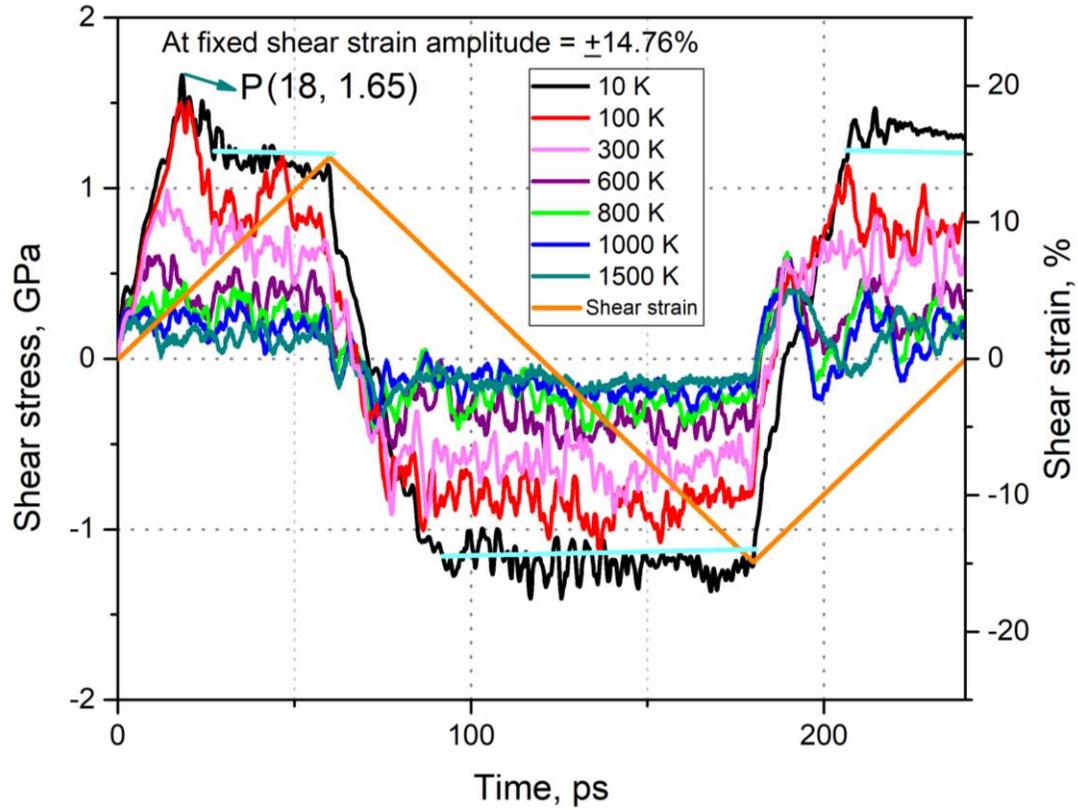

Fig. 9 Effect of temperature on the stress response of the system under applied cyclic shear strain amplitude of ±14.76%. LT cyan colored lines on the 10K curve indicate the averaged TMS values. Only first cycle values have been shown for brevity.

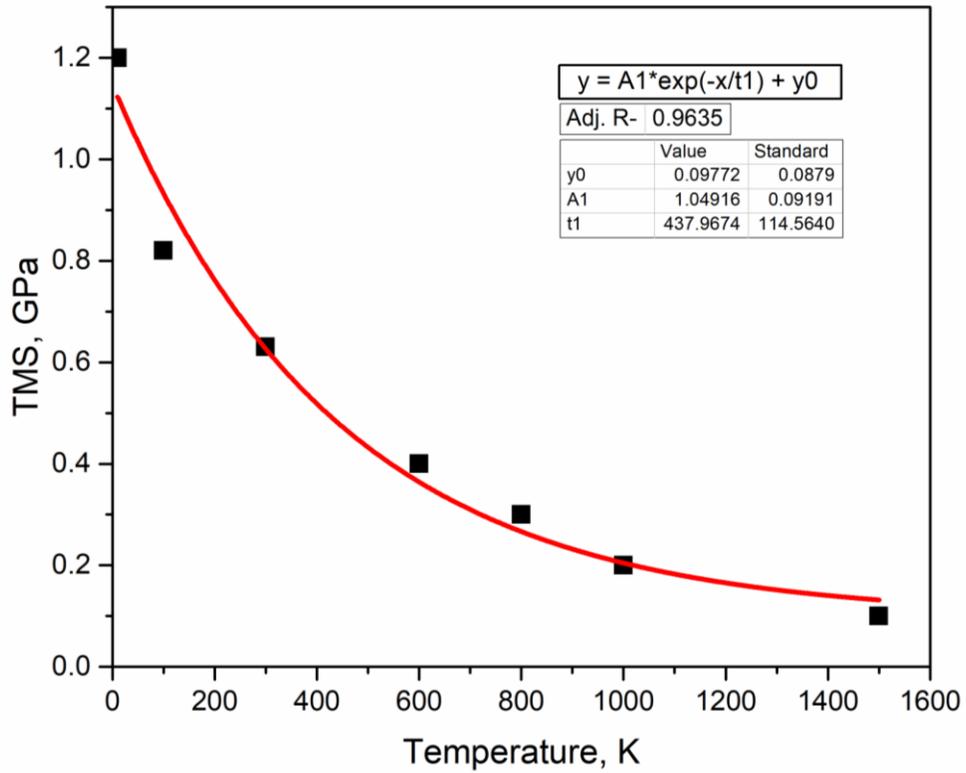

Fig. 10 An exponential decay nature of TMS value concerning temperature.

Stress-strain hysteresis loops at different temperatures are shown in Fig. 11. The area under the loop gives the resistive loss or damping per cycle in the material. It is evident that the loop area reduced with increasing temperature.

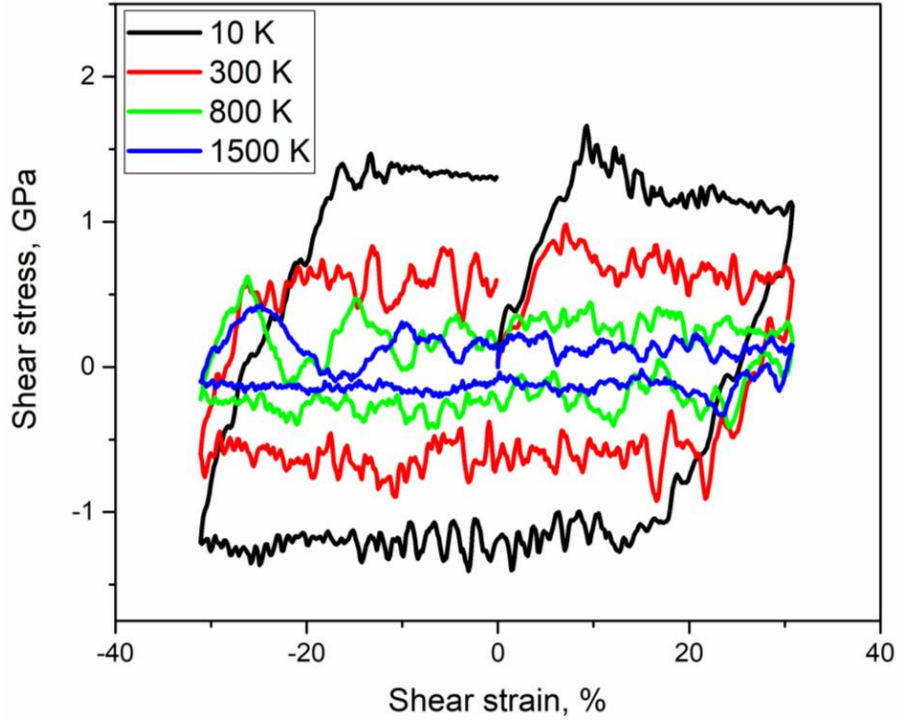

Fig. 11 Stress-strain hysteresis loops at different temperatures.

### 3.4 Effects of size and parallel twins

For further generalization, we studied twin migration at 10 K in large systems and with the presence of other twins. Firstly, shear stress was applied to large systems. The twin boundary migration was not observed once L/W ≥3; instead, deformation was initiated at the top rigid interface with the increase in shear loading. Offset value was increased with increasing size till L/W < 3. The effect of other parallel twins is shown in Fig. 12 (A-G) and Fig. 13 (A-G) for two different sizes, *viz*, $22.27 \times 6.05 \times 11.1$ nm$^3$ and $41.86 \times 6.05 \times 21.01$ nm$^3$ respectively, for a constant L/W of 2. In the smaller system (Fig. 12), all the parallel twins migrate. The middle twin was observed to move out of phase with respect to the other two surrounding twins in the forward and reverse applied stress. However, the middle twin moved only marginally (as indicated by blue arrows in Fig. 12) and has significant displacement only when the top twin approached close to it (Fig. 12D) and exerted image force. Interestingly, either the top or the bottom twin got almost complete reversibility at zero applied loads (Fig. 12C and E). One can also observe the asymmetry in the displacement of the twins during forward and reverse loading. Further cycling can modify these observations due to the appearance of image forces between TBs once they come closer. Fig.

12F and G represent the deformed nanowire after 10<sup>th</sup> and 50<sup>th</sup> cycles respectively. Thus, further application of cyclic loading modified the observations at the first cycle and led to the interaction between the twins. Further increase in size resulted in only the top twin's displacement, as shown in Fig. 13. Consistency of this can be observed during the progression of cycling as shown in Fig. 13F and G. However, the appearance of image forces could influence the behaviour at higher strain amplitudes. The authors also examined the possibility of TB migration when the applied shear stress was along the <110> direction and on another type of twist TB Σ3 {111} but have not noticed the phenomenon.

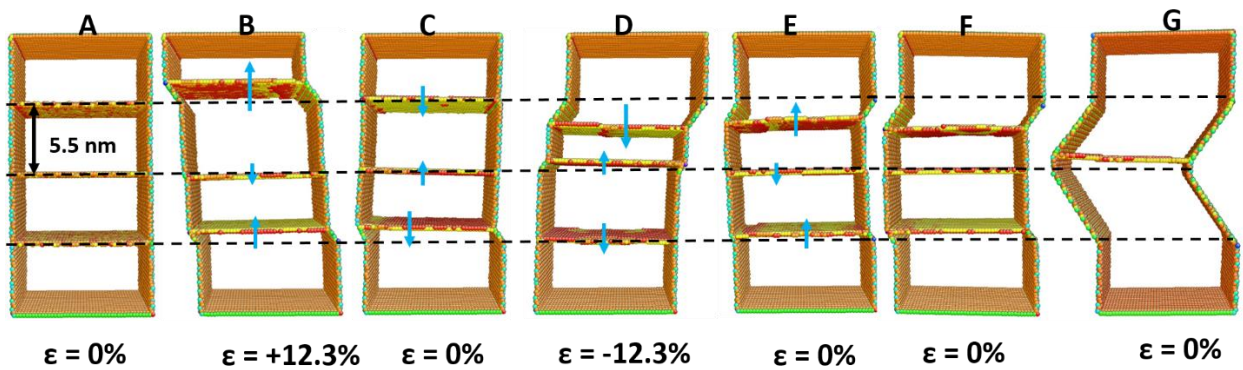

Fig. 12 Twin migration character in triple twin system. Green arrows indicate the migration or ready to migrate directions. Almost complete twin reversibility can be seen at C (for top twin), and at E (for middle and bottom twins). A-E indicate one complete cycle where F and G are deformed nanowires after completing the 10<sup>th</sup> and 50<sup>th</sup> cycles respectively.

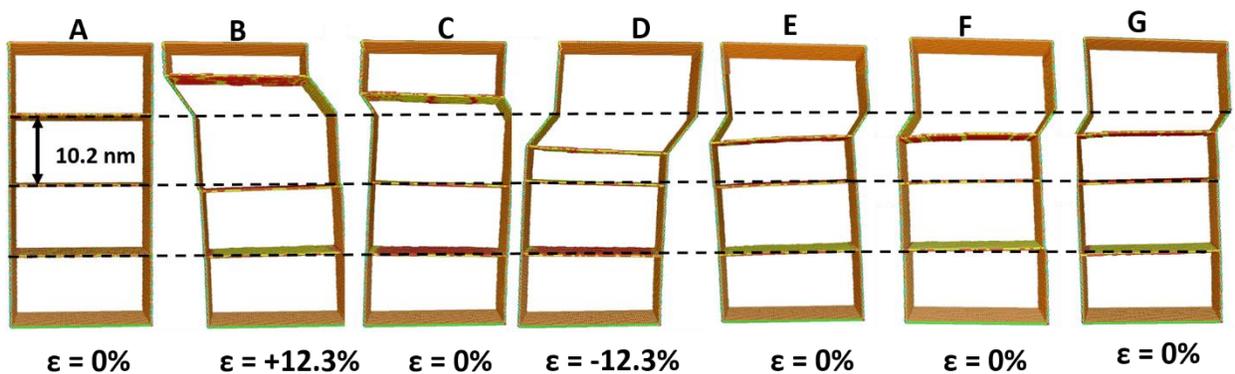

Fig. 13 Twin migration characters in the larger triple twin system. Only the top twin is migrating, whereas others are not. A-E indicate one complete cycle while F and G are deformed nanowires after completing the 10th and 50th cycles respectively.

### 3.5 Effects of boundary conditions and strain rates

We also investigated the effect of boundary conditions (BCs) on the above results. The motivation behind the surface boundary condition along Z-direction is to allow the creation of dislocations in a case from the surface since the burger vector is along Z-direction. This also allows the straightforward observation of SCM if presents. Periodic boundary condition (PBC) along Y-direction is not to restrict the motion of dislocations on the TB plane. Surface BCs along X-direction remove the repeatability of the TBs along X-direction. However, to check whether the irreversibility is not affected by the surface BC along Z-direction, we impose the periodicity along Z-direction and re-examine the above results.  This also allows the system to have the infinite TB slab which is helpful to study thin films.

PBC along Z removes the step formation as shown in Fig. 14. Apart from this, all the above results are consistent except with two significant changes. One was the reduction in offset value. For example, it reduced from around 11 (Fig. 5) to 5 $A^0$ at 10 K. This may be due to the absence of the friction offered by the surface along the plane in Z-direction. Further, this offset value increased with an increase in length for the fixed-width as shown in Fig. 14. The second one was the migration of the TB even when L/W = 6. However, this was subjected to the applied strain rate. Effects of strain rates along with the type of BC applied in Z-direction on the system are tabulated in table 3.

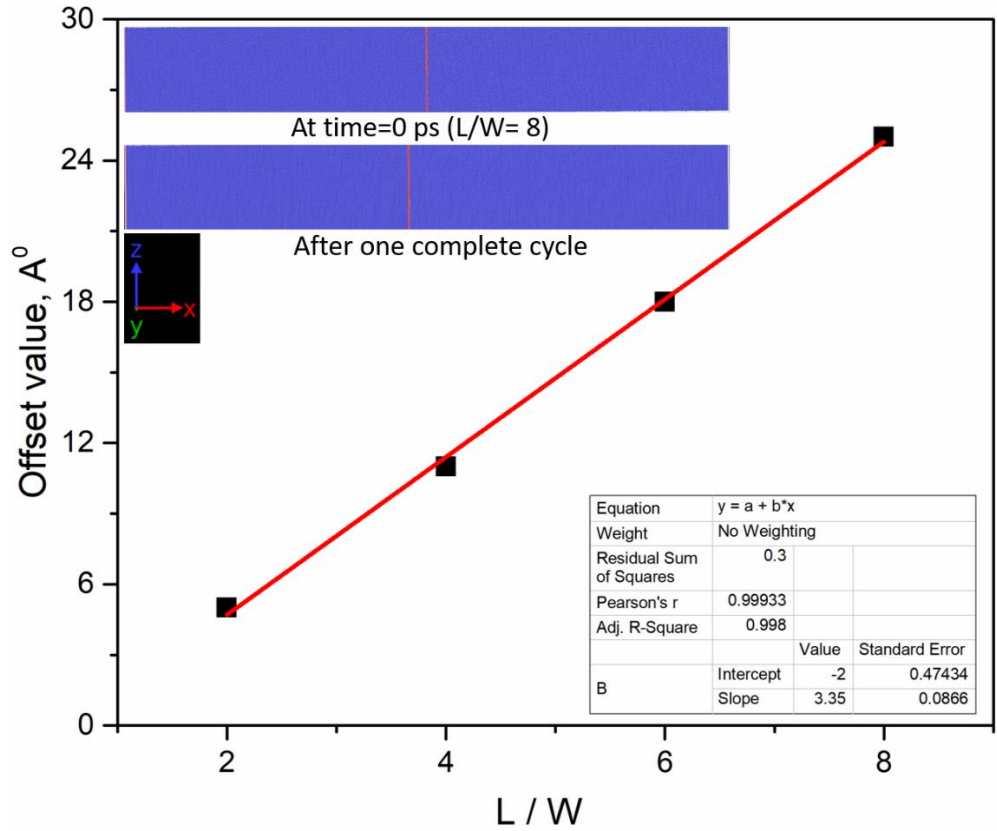

Fig. 14 Linear increase in offset value with increasing L/W ratio. Nanowires indicate the offset along with the absence of step formation once PBC is applied along Z-direction.

Table 3  Effect of strain rate on the system with the combination of the type of BC along Z-direction for the given periodic and surface BCs along with Y- and X- directions respectively.

| System size L/W = | Strain rate (s⁻¹) | Does PBC along Z? | Has TB migrated? |
|---|---|---|---|
| 2 | $5 \times 10^6$ | Yes | Yes |
| | | No | Yes |
| 2 | $5 \times 10^7$ | Yes | Yes |
| | | No | Yes |
| 2 | $5 \times 10^8$ | Yes | Yes |
| | | No | Yes |
| 2 | $5 \times 10^9$ | Yes | Yes, but deformation occurred after half a cycle |
| | | No | Yes, but deformation occurred after half a cycle |
| 6 | $5 \times 10^6$ | Yes | Yes |
| | | No | No |
| 6 | $5 \times 10^7$ | Yes | Yes |
| | | No | No |
| 6 | $5 \times 10^8$ | Yes | No |
| | | No | No |

| 6 | 5 x 10$^9$ | Yes | No |
| | | No | No |

## 4. DISCUSSION

From the above results, it can be understood that cyclic shear loading leads to irreversibility in twin boundary motion at lower temperatures. It is also observed that the magnitude of irreversibility is a weak function of the applied shear strain amplitude for a given temperature, generally showing a saturation with increasing strain (Fig. 5). This indicates that the twin boundary irreversibility is due to the inertia of the twin plane. The origin of the inertia can be understood from the resistance offered by the system and threshold stress (opposite and equivalent to the minimum friction stress offered by the system at a given temperature) on the twin plane seems necessary for it to move or reverse the direction (Fig. 4 and 9). Stresses developed on the twin plane upon reversing the strain at B were not identical to the situation at D (Fig. 2) because of the slipping step lying in between the moving top rigid portion and the twin in the latter case. This caused the asymmetry in the $t_{lag}$ for the forward and reverse cases in Fig. 4. This can be further confirmed from the equal $t_{lag}$ for the forward and reverse cases in Fig. 15 for the same size of the system which was represented in Fig. 4. This was due to the absence of the step formation. However, there was a slight decrease and increase in $t_{lag}$ at zero strain and peak strains respectively in Fig. 15 compared to Fig. 4. This was due to the inertia of the TB in the absence of surface resistance. The reduction in stress from location $P$ in Fig. 9 is due to the static friction in the system. Dynamic friction prevails in the system once the twin plane starts moving, leading to identical TMS at each peak strain. Further, the fluctuations in the shear stress around the TMS value (Fig. 9) are due to the layer-by-layer propagation of the twin plane [35,36]. Fluctuations in displacement at higher temperatures, as shown in Fig. 6 can be attributed to the wavy nature of the twin plane at higher temperatures [37, 38]. As shown in Fig. 9, stresses developed in the system for fixed strain amplitude reduced with an increase in temperature. This may indicate that thermal fluctuations in the system enhance the formation and displacement dynamics of the edge partials by reducing the friction in the system at higher temperatures. Earlier, Zhou et al. [36] reported a temperature-induced reduction in critical shear stress (or TMS) for twin boundary migration in fcc metals. Atomic fluctuations can favor the decrease in free energy across the boundary to cause easy motion of the boundary. This resulted in exponential decay in the TMS value with increasing temperature, as shown in Fig. 10. The geometric coupling factor analysis further explained the reduction in sliding resistance offered by the system with increasing temperature reflected as the drop in the damping loss (Fig. 11) and the TMS value.

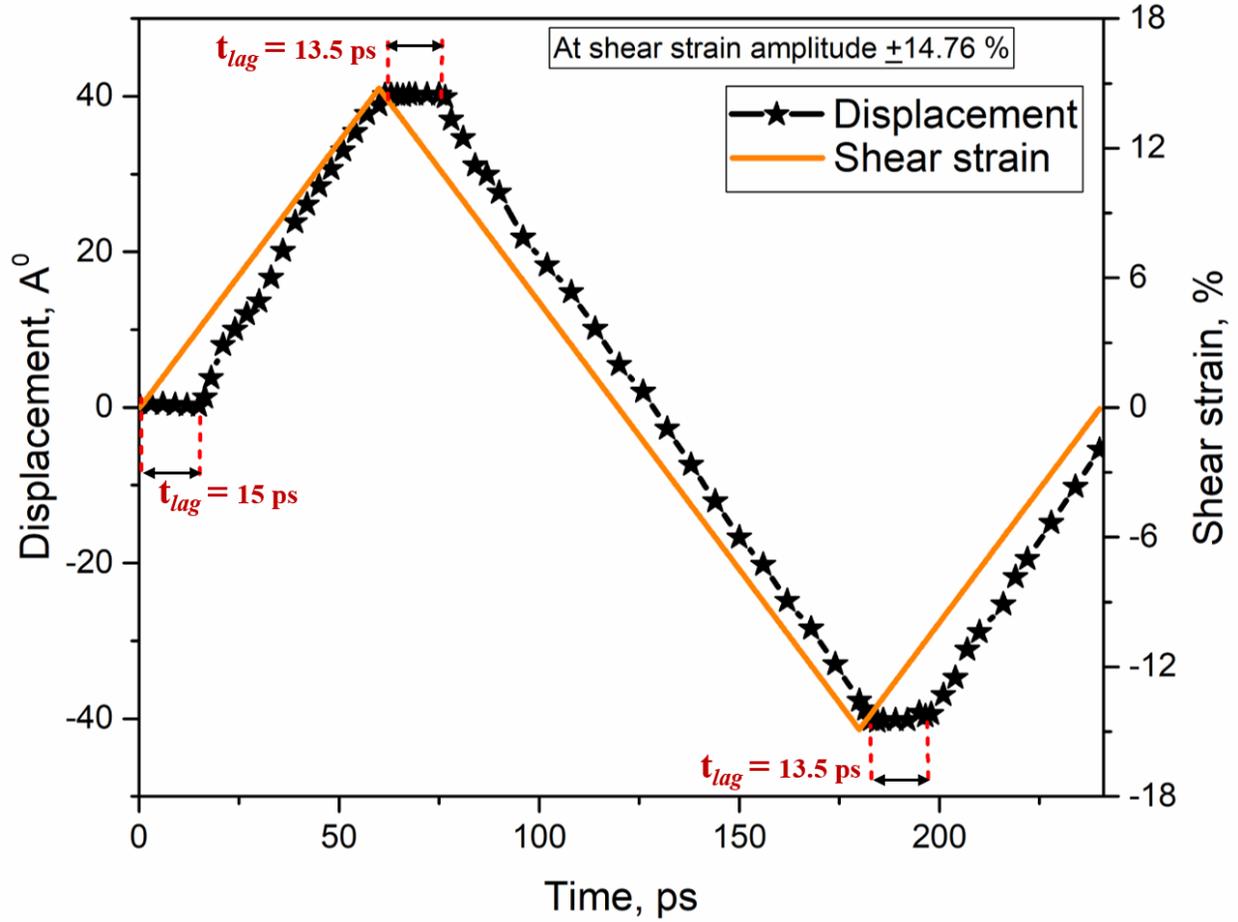

Fig. 15 Cyclic nature of twin boundary motion at 10 K for a shear strain amplitude of ±14.76% when PBC is applied along Z- direction.

For checking the consistency of the results on further cyclic shear loading, simulations were performed up to 50 cycles for all the test conditions mentioned in table 1. Results were stabilized from the 1st cycle onwards at each test condition. For brevity, cyclic stress response for the first 50 cycles is shown in Fig. 16 at the shear strain amplitude ±14.76 % and temperature 600 K. However, the presence of other twins or the increase in system sizes can alter the results. The presence of other twins may lead to a redistribution of stress on surrounding twins upon their migration and impose image forces. At the larger sizes, the same strain can cause more considerable bending moment on the system. This further results in normal stresses on the twin planes [9]. Thus, large normal stresses can resist the twin migration. However, cyclic TB migration in the large systems could be evident if PBC is applied in Z- direction along with the strain rates below $\sim 10^8$ /s. The nanowires were deformed via the initiation of twins from the fixed rigid portion along with or without the TB migration at the higher strain rates during the progression of cycling. This may be due to the inertia associated with the planes along the X-direction. $v_n$ was reduced with increasing the system size. For example, it decreased from 0.85 to 0.21 A$^0$/ps when the system size was increased from L/W=2 to 6 respectively for the simulation conditions were given in S.N. 3 in table 1 with the PBC along Z-direction. This caused the increase in offset value with an increase in system size (Fig. 14). This is probably due to the less stress distribution at the TB when the length has increased.

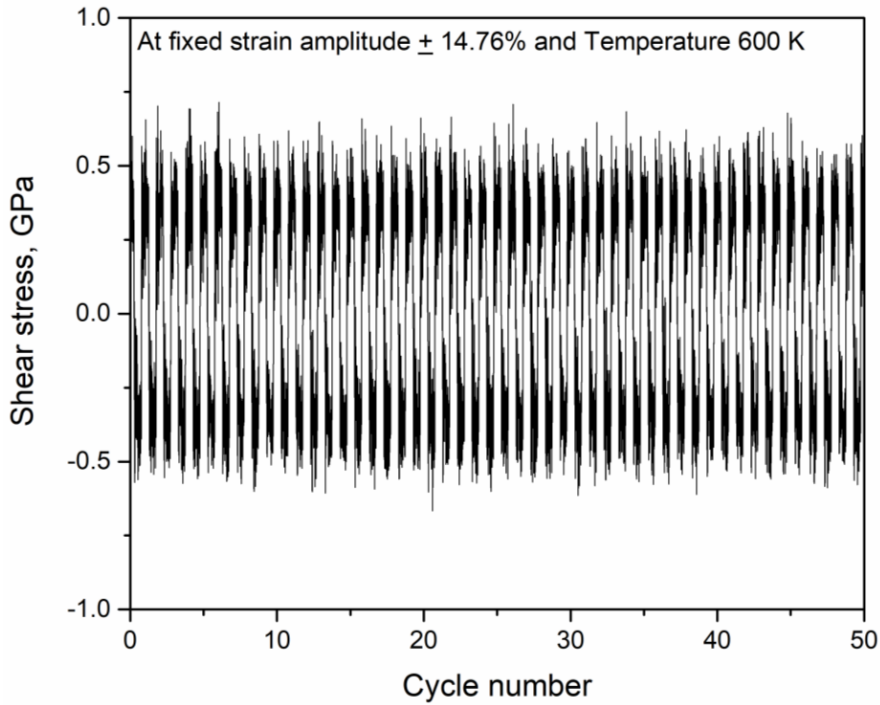

Fig. 16 Cyclic shear stress response of the system for the first 50 cycles at given shear strain amplitude of ±14.76% and temperature at 600 K.

**Conclusions**

We studied the cyclic irreversibility characteristics of twin boundaries in BCC-Fe under shear fatigue conditions. Twin boundary migrated step by step under cyclic shear load. Shear strain amplitude did not significantly influence the twin boundary offset, while temperature strongly influenced the same. TB exhibited cyclic irreversibility character at lower temperatures, but near reversibility was observed at 1500 K. An increase in sliding velocity and a decrease in normal velocity have been observed with increased temperature. Exponential reduction in the offset and TMS values was identified with temperature. The mechanism governing the displacement of the twin boundary was identified as the activity of 1/6 ⟨111⟩ = 1/12 ⟨111⟩ + 1/12 ⟨111⟩ type edge partial dislocations on adjacent {112} twin boundaries. Cyclic shear fatigue results were stabilized from $1^{st}$ cycle onwards. Almost complete twin reversibility was observed in the triple parallel-twin system. However, the presence of image forces at larger strains removed this unison in further cycling. Twin migration was seen to be resisted by the normal stresses developed on twin planes at larger system sizes. However, PBC in Z- direction and the strain rates below $\sim 10^8$ /s led the TB to migrate at the larger nanowires. Offset value linearly increased with the increasing length of the nanowire.

**Acknowledgments**

The authors are thankful to Dr. A.K. Bhaduri, Director, IGCAR, Dr. Shaju K. Albert, Director, Metallurgy and Materials Group, IGCAR, Dr. R. Divakar, Associate director, MMG, IGCAR and Dr. M. Vasudevan, Head, MDTD, IGCAR, Kalpakkam for their keen interest and support in this investigation.

**Data availability**

The raw/processed data required to reproduce these findings cannot be shared at this time as the data also forms part of an ongoing study.